\documentstyle[12pt]{article}
\begin{document}
\newcommand{\Si}{\Sigma}
\newcommand{\tr}{{\rm tr}}
\newcommand{\h}{{\hbar}}
\newcommand{\ad}{{\rm ad}}
\newcommand{\Ad}{{\rm Ad}}
\newcommand{\ti}[1]{\tilde{#1}}
\newcommand{\om}{\omega}
\newcommand{\Om}{\Omega}
\newcommand{\de}{\delta}
\newcommand{\al}{\alpha}
\newcommand{\te}{\theta}
\newcommand{\vth}{\vartheta}
\newcommand{\be}{\beta}
\newcommand{\pa}{\partial}
\newcommand{\la}{\lambda}
\newcommand{\La}{\Lambda}
\newcommand{\D}{\Delta}
\newcommand{\ve}{\varepsilon}
\newcommand{\ep}{\epsilon}
\newcommand{\vf}{\varphi}
\newcommand{\Ph}{\Phi}
\newcommand{\G}{\Gamma}
\newcommand{\ka}{\kappa}
\newcommand{\ip}{\hat{\upsilon}}
\newcommand{\Ip}{\hat{\Upsilon}}
\newcommand{\ga}{\gamma}
\newcommand{\ze}{\zeta}
\newcommand{\si}{\sigma}
\renewcommand{\thefootnote}{\alph{footnote}}
\def\bfa{{\bf a}}
\def\bfb{{\bf b}}
\def\bfc{{\bf c}}
\def\bfd{{\bf d}}
\def\bfm{{\bf m}}
\def\bfn{{\bf n}}
\def\bfp{{\bf p}}
\def\bfu{{\bf u}}
\def\bfv{{\bf v}}
\def\bft{{\bf t}}
\def\bfx{{\bf x}}
\newcommand{\li}{\lim_{n\rightarrow \infty}}
\newcommand{\mat}[4]{\left(\begin{array}{cc}{#1}&{#2}\\{#3}&{#4}
\end{array}\right)}
\newcommand{\beq}[1]{\begin{equation}\label{#1}}
\newcommand{\eq}{\end{equation}}
\newcommand{\beqn}[1]{\begin{eqnarray}\label{#1}}
\newcommand{\eqn}{\end{eqnarray}}
\newcommand{\p}{\partial}
\newcommand{\di}{{\rm diag}}
\newcommand{\oh}{\frac{1}{2}}
\newcommand{\su}{{\bf su_2}}
\newcommand{\uo}{{\bf u_1}}
\newcommand{\uu}{{\bf u_2}}
\newcommand{\GL}{{\rm GL}(N,{\bf C})}
\newcommand{\SL}{{\rm SL}(N,{\bf C})}
\def\sln{{\rm sl}(N,{\bf C})}
\newcommand{\gl}{gl(N,{\bf C})}
\newcommand{\PSL}{{\rm PSL}_2({\bf Z})}
\def\f1#1{\frac{1}{#1}}
\newcommand{\rar}{\rightarrow}
\newcommand{\upar}{\uparrow}
\newcommand{\sm}{\setminus}
\newcommand{\ms}{\mapsto}
\newcommand{\bp}{\bar{\partial}}
\newcommand{\bz}{\bar{z}}
\newcommand{\bA}{\bar{A}}
\hyphenation{Calo-gero-Moser}
\vspace{0.3in}
\begin{flushright}
 ITEP-TH-7/01\\
hep-th/0102167\\
\end{flushright}
\vspace{10mm}
\begin{center}
{\Large\bf Sklyanin Bracket and Deformation of the Calogero-Moser System}\\[1cm]
V.A. Dolgushev\\[0.5cm]
{\it On leave in Tomsk State University, Tomsk, Russia;\\
Bogoliubov Laboratory of Theoretical Physics, \\
JINR, Dubna, Moscow Reg., 141 980, Russia;\\
ITEP, 117259 Moscow, Russia;\\
Address: Tomsk, 634050, Lenin ave. 36, Tomsk State University, Physics
Department\\
Phone: +7-3822-426243 Fax: +7-095-8839601 \\
E-mail: vald@thsun1.jinr.ru, dolgushev@gate.itep.ru}\\[0.5cm]
\end{center}
\begin{abstract}
A two-dimensional integrable system being a deformation of the rational Calogero-Moser
system is constructed via the symplectic reduction,
performed with respect to the Sklyanin algebra action.
We explicitly resolve the respective classical equations of motion via the projection
method and quantize the system.
\end{abstract}
~\\[0.5cm]
\pagestyle{headings}
\section{Introduction}
The relationship between the integrable systems and the Lie group gauge
symmetries is well known \cite{OPclass},\cite{OPquant},\cite{Hit},\cite{Nikita}.
However, Lie groups
by no means exhaust all the symmetries and in fact there are a lot of
systems whose gauge symmetries cannot be reduced to the Lie group action
\cite{Hen}\,. For this reason it is quite interesting to construct an integrable
system using the non Lie group symmetries.
The aim of the paper is to present the example of such integrable
system.
We obtain the system via the symplectic
reduction from an extended phase space, where the symmetries are realized
by the action of the quadratic Sklyanin algebra \cite{Skl1}. Let us
briefly explain the key idea for realizing this action.
Note that the well known
coadjoint action of a Lie algebra on its dual space can be written in
terms of the linear Poisson structure, naturally associated to the Lie algebra.
Namely if $p_{\mu}$ are coordinates in the dual space and the respective Poisson
tensor is of the form
$$
\al_{\mu\nu}(p)=f^{\la}_{\mu\nu}p_{\la},
$$
where $f^{\la}_{\mu\nu}$ are the structure constants of the Lie algebra, then the
coadjoint action is rewritten as
\begin{equation}
\de_{\ep}p_{\mu}=\al_{\mu\nu}(p)\ep^{\nu}(p),
\label{coad}
\end{equation}
where $\ep^{\nu}$ are the infinitesimal parameters of the action.
If we assume now that the Poisson tensor $\al_{\mu\nu}(p)$ in
(\ref{coad}) is no longer
linear in $p_{\mu}$ we get the generalization of the coadjoint action to
the case of an arbitrary Poisson manifold. Such a generalization of the
coadjoint action has been originally proposed by Karasev \cite{Ka} and in general
it cannot be reduced to the Lie algebra action in a sense that one cannot choose the
global basis of the sections $\ep^{\mu}(p)$ with a Lie algebra commutation relations of
the transformations (\ref{coad}). Instead the form of the commutation relations in a general case
looks as follows
$$
[\de_{\xi},\de_{\eta}]=\de_{\ep},
$$
where
\begin{equation}
\begin{array}{c}
\ep = [\xi ,\eta]= d \al(\xi,\eta)+\al(d\xi,\eta)+
\al(\xi,d\eta),\\
~\\
\ep^{\mu}=\pa^{\mu}(\al_{\nu\la}\xi^\nu \eta^{\la})
 +
\al_{\nu\la}
(\pa^{\nu}\xi^{\mu} - \pa^{\mu}\xi^{\nu})
\eta^{\la} +
\al_{\nu\la}\xi^{\nu}
(\pa^{\la}\eta^{\mu} - \pa^{\mu}\eta^{\la}),
\end{array}
\label{comm}
\end{equation}
$$
\pa^{\mu}=\frac{\pa}{\pa p_{\mu}}.
$$
Thus we get an example of the manifold, equipped with the non Lie
algebra symmetries.
A mathematical rigor treatment of such symmetries involves a definition of
the so-called Lie algebroid \cite{Mcz},\cite{H-M}, which is defined as a bundle $\cal A$ over a
manifold $M$ with a Lie bracket, given in the set of sections $\Gamma({\cal A})$
and an anchor map $\de~:~\Gamma({\cal A}) \mapsto \Gamma (TM)$\,, being a
homomorphism of the respective Lie algebras.
Using this terminology we see that the above construction is an example of
the Lie algebroid. Namely, the Lie algebroid is the
cotangent bundle over the Poisson manifold. The Lie bracket between
the sections $\xi$ and $\eta$ is given by the equation (\ref{comm}) and
the anchor map is defined by the formula
(\ref{coad})\footnote{such example of the Lie algebroid, naturally associated to a Poisson
manifold has been originally presented in the paper \cite{M-Xu}.}.
Note that the Jacobi identity of the Lie bracket (\ref{comm})
holds in virtue of the respective Jacobi identity for the Poisson tensor
$\al_{\mu\nu}(p)$\,.
In this paper we start from the Poisson manifold ${\bf R}^4$, equipped with the
standard classical Sklyanin bracket \cite{Skl1}\,, and thus equipped with the
algebroid action, which may be regarded as the action of the Sklyanin
algebra. We choose the extended phase space of the integrable system to be
the cotangent bundle over ${\bf R}^4$ with the standard symplectic
structure. We lift the algebroid transformations
from the base manifold to the canonical ones in the cotangent bundle and
readily find the respective Hamiltonian generators, which define a natural
analogue of the momentum map for the case of the non Lie group symmetries.
Then we construct the phase space of the integrable system by
performing the symplectic reduction on the surface
of the non zero level of the generators. The underlying
Hamiltonians are obtained by reduction of the Casimir functions of the Sklyanin
bracket.
Note that since the Sklyanin bracket \cite{Skl1} is the deformation of the
Poisson bracket, which is equivalent to the linear bracket, associated
to the Lie algebra $\uu$ the presented symplectic reduction
is similar to the one leading to the rational Calogero-Moser
system \cite{OPclass},\cite{CM}. In particular, one of the exponential degenerations
of the Sklyanin bracket leads to the system, whose dynamics is in a sense
equivalent to the dynamics of the two-particle rational Calogero-Moser
system.
As an aside, it is worth noting that although the integrable system is
obtained via the symplectic reduction we still cannot find the respective Lax
representation because the symmetries, that have been used for
constructing the system are not Lie group ones. However, we show that it is possible to
resolve the equations of motion for our system via the projection method.
So the problem of finding the respective Lax representation
is seemed to be rather intriguing and we are going to comment on it in the concluding section.
The organization of the paper is as follows. In the second section we
present the construction of the integrable system. In the third section we
develop the projection method and explicitly resolve the equations of
motion. The fourth section is devoted to the quantization. We find that
although it is not possible to pose the bound states problem for our system
one of the Hamiltonians possesses a discrete spectrum. At last, in the
concluding section we discuss some open questions.
All through the paper the Greek indices take values from $0$ to $3$
the Latin indices run from $1$ to $3$ and whenever the indices $i,j,k$
are
meat they are assumed to take values $1,2,3$ together with their cyclic
permutations. $\pa^{\mu}$ denotes
$\displaystyle \frac{\pa}{\pa
p_{\mu}}$ and $\pa_{\nu}$ is reserved for $\displaystyle \frac{\pa}{\pa
x^{\nu}}$\,.
\section{Construction of the System}
We construct the system starting with the extended phase space, which is
chosen to be the cotangent bundle $T^*{\bf R}^4$\,, equipped with the
canonical symplectic structure
\begin{equation}
\om=dp_{\mu}\wedge dx^{\mu}.
\label{symp}
\end{equation}
Here we use the following unusual conventions. The coordinates
on the base ${\bf R}^4$ are denoted by
$p_{\mu}$ and the coordinates in the cotangent
space are denoted by $x^{\mu}$\,.
Let us endow the base manifold ${\bf R}^4$ with the
quadratic Sklyanin bracket
\begin{equation}
\{p_0,p_i\}=\al_{0i}(p) =J_{jk}p_j p_k
\qquad
\{p_i,p_j\}=\al_{ij}(p) = p_0 p_k, \label{Skl}
\end{equation}
where the coefficients $J_{12}$, $J_{23}$ and $J_{31}$ satisfy
the identity
\begin{equation}
J_{12}+J_{23}+J_{31}=0,
\label{cycle}
\end{equation}
which is equivalent to the Jacobi identity for the
bracket (\ref{Skl}). The condition (\ref{cycle}) also implies
that these coefficients can be represented in the form
\begin{equation}
J_{ij}=J_i-J_j,
\label{form}
\end{equation}
where the values $J_i$ are defined up to
the additional constant $c$, $J_i\,\mapsto\, J_i+c$\,.
Following the steps, outlined in the introduction we firstly define the
analogue of the coadjoint transformations on the base manifold
${\bf R}^4$ using the Poisson structure (\ref{Skl}).
Namely, the infinitesimal coadjoint transformations are defined as
\begin{equation}
\de_{\ep} p_{\mu}=\al_{\mu\nu}(p)\ep^{\nu}(p),
\label{ancor}
\end{equation}
where $\ep^{\nu}$ are the respective infinitesimal parameters.
The transformations (\ref{ancor}) can be easily lifted to the
canonical ones in the cotangent bundle
\begin{equation}
\de_{\ep} p_{\mu}=\al_{\mu\nu}\ep^{\nu}(p)
\qquad
\de_{\ep} x^{\mu} = \pa^{\mu} (\al_{\nu\la} \ep^{\nu}(p)x^{\la})
\label{transf}
\end{equation}
and the Hamiltonian generators
corresponding to the transformations (\ref{transf}) can be also
readily found\footnote{thus we have
constructed the so-called Hamiltonian algebroid \cite{LO} over $T^*{\bf
R}^4$\,; in the paper \cite{LO} it is shown that the
natural construction of the Hamiltonian algebroid exists for
an arbitrary Lie algebroid}
\begin{equation}
M_{\ep}=\al(\ep,x)=\al_{\mu\nu}(p)\ep^{\mu}x^{\nu}.
\label{gen}
\end{equation}
Then in order to obtain a non-trivial dynamical system via the
symplectic reduction method we consider the surface of
non-zero level of the generators (\ref{gen}) in  the extended phase
space $T^*{\bf R}^4$.
In analogy to what it is done for case of the rational Calogero-Moser
system we set
\begin{equation}
\begin{array}{c}
M_0=J_{23}p_2p_3 x^1+J_{31}p_3p_1 x^2+J_{12}p_1p_2 x^3=0 \\
~\\
M_1=p_0(x^2 p_3 - x^3 p_2) - J_{23}p_2 p_3 x^0 = \nu \\
~\\
M_2=p_0(x^3 p_1 - x^1 p_3) - J_{31}p_3 p_1 x^0 = 0 \\
~\\
M_3=p_0(x^1 p_2 - x^2 p_1) - J_{12}p_1 p_2 x^0 = 0.
\end{array}
\label{constr}
\end{equation}
The symplectic form (\ref{symp}) being reduced on the constraint
surface (\ref{constr}) is of course degenerate and the respective
kernel distribution is given by the following vector field, being
tangent to the surface (\ref{constr})
\begin{equation}
\begin{array}{cc}
\de p_0 =J_{23}p_2 p_3  &   \de x^0 = (x^2 p_3 -x^3 p_2) \\
~\\
\de p_1 = 0                &     \de x^1 = 0 \\
~\\
\de p_2 = -p_0 p_3   &    \de x^2 = -(p_0 x^3+J_{23} p_3
x^0)\ep^1 \\
~\\
\de p_3 = p_0 p_2    &   \de x^3 = (p_0 x^2-J_{23} p_2 x^0).
\end{array}
\label{ker}
\end{equation}
Our system is then presented as a result of the symplectic reduction on the
constraint surface (\ref{constr}) with respect to the transformations,
corresponding to the vector field (\ref{ker}). In what follows, these
transformations are called the gauge ones and the integral curves of the
vector field (\ref{ker}) are called the gauge orbits.
We perform the symplectic reduction using the gauge fixing condition
$x^2=0$. In fact it is easy to see that this gauge condition
has a horizon, given by the equation
\begin{equation}
p_0 x^3-J_{32} p_3 x^0=0 \,.
\label{hor}
\end{equation}
Namely, for the points, satisfying (\ref{hor})
the coordinate $x^2$ is unchange under the
gauge transformations (\ref{ker}), and hence for
these points the value of the coordinate $x^2$
in general cannot be reduced to zero with the help of the gauge
transformations. However, in the paper we do not bother
the subtlety and we are not going to
discuss it.
So taking into account the condition $x^2=0$ together with the
constraint equations (\ref{constr}) we get the embedding of the reduced
phase space into the extended one. Namely, the embedding is given by
the following mapping
\begin{equation}
\begin{array}{cccc}
\displaystyle
x^0=u^1 & x^1=0 & x^2=0 & x^3= u^2 \\[0.2cm]
p_0=v_1 & p_1=0 &\displaystyle p_2=-\frac{\nu}{v_1 u^2 - J_{32} v_2 u^1}
&
p_3=v_2,
\end{array}
\label{emb}
\end{equation}
where $u^1, v_1, u^2, v_2$ are the canonical coordinates parameterizing
the reduced phase space.
The pair of the Casimir functions of the Sklyanin bracket
\begin{equation}
C_1=\frac{1}{2}\sum_{l} (p_l)^2 \qquad
C_2=\frac{1}{2}((p_0)^2+\sum_{l} J_l (p_l)^2) \label{Cas}
\end{equation}
provides us with the required set of the commuting and independent
Hamiltonians of the presented system. Really, the gauge invariance of
the Casimir functions and their involution property with respect to the
canonical structure on $T^*{\bf R}^4$ ensures that the Hamiltonians,
obtained by reduction of (\ref{Cas}) on the surface of the constraint equations and
of the gauge fixing condition are commuting and independent.
Thus we readily find the Hamiltonians for the our system
\begin{equation}
H_1=\frac{(v_2)^2}{2}+\frac{\nu^2}{2(v_1 u^2 - J_{32} u^1 v_2)^2}
\qquad
H_2=\frac{(v_1)^2}{2}+J_{32}\frac{(v_2)^2}{2},
\label{rham}
\end{equation}
where we set $J_2=0$ using the arbitrariness of the coefficients
$J_i$\,.
We have presented the example of the integrable system, obtained via the
symplectic reduction, which is performed in the context of the non Lie group
symmetries. The most similar symplectic reduction performed in the group
theoretical context is the one leading to the two-particle rational Calogero-Moser
system \cite{OPclass},\cite{CM}. This resemblance of the reductions becomes quite
clear if we compare the Poisson manifold ${\bf R}^4$ (\ref{Skl})
to the dual space of the Lie algebra $\uu$ and the coadjoint transformations (\ref{ancor})
to the coadjoint action of the Lie algebra $\uu$\,.
Note that the exponential degeneration of the Sklyanin bracket
corresponding
to case $J_{23}=0$ leads us to the system with the following
Hamiltonians
\begin{equation}
H_1=\frac{(v_2)^2}{2}+\frac{\nu^2}{2(v_1 u^2)^2}
\qquad
H_2=\frac{(v_1)^2}{2},
\label{=CM}
\end{equation}
which describe the dynamics of the two-particle Calogero-Moser system
with the coupling constant, depending on the initial value of $v_1$ and with
$u^1$\,, being the coordinate of the mass center.
\section{Projection Method}
In the this section we show that in spite of the more complicated nature
of
the gauge transformations  the equations of motion for the presented
system
(\ref{rham}) can be explicitly resolved via the projection method.
Let us briefly recall the main idea of the method. It is well known that
each point of the reduced phase space corresponds to a gauge orbit on
the constraint surface, where the dynamics is described by the gauge
invariant
Hamiltonians ((\ref{Cas}) in our case). We integrate the equations of
motion defined by the Hamiltonians (\ref{Cas}) and construct
the surface, which consists of the gauge orbits,  passing through to all the
points of the evolution curves. The intersections of the surface with the level
of the gauge fixing condition are just the evolution curves of the reduced
system. Thus the explicit solutions of the evolution equations for the
reduced system can be derived if the explicit description of the gauge
orbits is known.
In order to describe the gauge orbits we have to integrate the system of
differential equations given by the kernel vector field (\ref{ker})
\begin{equation}
\begin{array}{cc}
\displaystyle
\frac{d}{ds} p_0 =J_{23}p_2 p_3  &
\displaystyle
  \frac{d}{ds} x^0 = (x^2 p_3 -x^3 p_2)  \\
~\\
\displaystyle
\frac{d}{ds} p_1 = 0                & \displaystyle    \frac{d}{ds} x^1
= 0 \\
~\\
\displaystyle
\frac{d}{ds} p_2 = -p_0 p_3   &  \displaystyle  \frac{d}{ds} x^2 =
-(p_0 x^3+J_{23} p_3
x^0) \\
~\\
\displaystyle
\frac{d}{ds} p_3 = p_0 p_2   & \displaystyle \frac{d}{ds} x^3 =
(p_0 x^2-J_{23} p_2
x^0),
\end{array}
\label{flow}
\end{equation}
where $s$ parameterizes the gauge orbit. Note that the r.h.s. of the
differential equations for the momenta contains the momenta only. Hence
we can integrate these equations separately. Then using the fact that the
gauge
transformations (\ref{ker}) are canonical ones we can resolve the rest
equations for the coordinates with the help of the following formula
\begin{equation}
x^{\mu}(s)=\frac{\pa p^0_{\nu}}{\pa p_{\mu}} x_0^{\nu},
\label{coor}
\end{equation}
where  $x^{\mu}_0, p^0_{\nu}$ denote the initial conditions for the
system
(\ref{flow}). Note also that the phase coordinates $x^1,~p_1$ are not
changed under the gauge transformations so in what follows we are going
to omit them.
It is easy to see that the system of differential equation for the
momenta
is similar to the system of equations for the ordinary ${\bf so_3}$ top
\cite{Audin}. Using the analogy we can readily find two independent
integrals of motion for the system (\ref{flow})
\begin{equation}
A=\sqrt{(p_2)^2+ (p_3)^2} \qquad
B=\sqrt{(p_0)^2 + J(p_3)^2},
\label{AB}
\end{equation}
where $J=J_{32}$ (in the rest of the paper we assume that $J>0$).
Let us consider the integrals (\ref{AB}) as the new independent momenta.
Then the rest third coordinate in the momentum space can be chosen in
two different ways. First, one can introduce the following angle variable
\begin{equation}
\vf=arctg (p_3/p_2),
\label{ang1}
\end{equation}
for which the momenta $p_0,p_2,p_3$ take the following form
\begin{equation}
p_0=\sqrt{B^2-JA^2sin^2\vf} \qquad
p_2=A cos \vf \qquad
p_3=A sin \vf
\label{<1}
\end{equation}
Second, the third coordinate can be chosen as
\begin{equation}
\vf_2=arctg (\sqrt{J}p_3/p_0).
\label{ang2}
\end{equation}
And in this case we get
\begin{equation}
p_0=B cos \vf_2 \qquad
p_2=\sqrt{A^2-\frac{B^2}{J}sin^2\vf_2}  \qquad
p_3=\frac{B}{\sqrt{J}} sin \vf_2
\label{<2}
\end{equation}
The first way to choose the new momenta is applicable when
\begin{equation}
k=\frac{\sqrt{J}A}{B}<1,
\label{ine}
\end{equation}
because it is the case when the expression $B^2-JA^2sin^2\vf$ do not
change the sign. Otherwise, one should use the second way of choosing
coordinates in the momentum space.
In both cases one can develop the projection method and find the
evolution
equations of the reduced system proceeding similarly at each stage. So in
our paper we are going to discuss the projection method only for the
first case
(\ref{ine}) omitting another possibility.
In what follows, it is more convenient to use instead of $x^0,x^2,x^3$
the
coordinates, which are canonically conjugated to $A,B$ and $\vf$. They
can be
readily found in the following form
\begin{equation}
\begin{array}{l}
\displaystyle
x_A= x^2 cos \vf + x^3 sin \vf -
\frac{JA sin^2 \vf}{\sqrt {B^2-J A^2 sin^2 \vf}}x^0 \\[0.5cm]
\displaystyle
x_B=\frac{B x^0}{\sqrt {B^2-J A^2 sin^2 \vf}} \\[0.5cm]
\displaystyle
x_{\vf}= - A x^2 sin\vf  + A x^3 cos\vf
-\frac{JA^2 sin \vf cos \vf}{\sqrt {B^2-J A^2 sin^2 \vf}} x^0.
\end{array}
\label{newc}
\end{equation}
In terms of $A,B,\vf,x_A, x_B$ and $x_{\vf}$ the solution for the system
(\ref{flow}) looks as follows
\begin{equation}
\begin{array}{c}
\vf(s)=am(Bs+F(\vf_0;k);k) \qquad
A(s)=A_0 \qquad B(s)=B_0 \\[0.5cm]
\displaystyle
x_{\vf}(s)=\sqrt{\frac{1-k^2 sin^2 \vf_0}{1-k^2 sin^2\vf}}
x^0_{\vf}\\[0.5cm]
\displaystyle
x_A(s)=x^0_A+\frac{\sqrt{J}}{B}\sqrt{1-k^2 sin^2 \vf_0}\,
(F_k(\vf;k)-F_k(\vf_0;k))x^0_{\vf} \\[0.5cm]
\displaystyle
x_B(s)=x^0_B-\sqrt{1 - k^2 sin^2 \vf_0}\,(s+
\frac{A\sqrt{J}}{B^2}[F_k(\vf;k)-F_k(\vf_0;k)])x^0_{\vf},
\label{ch}
\end{array}
\end{equation}
where $u=am(t;k)$ is the elliptic Jacobi function being
the inverse function to the first order elliptic integral
$$
F(u;k)=\int_0^u\frac{dt}{\sqrt{1-k^2 sin^2 t}},
$$
$F_k(u;k)$ denotes the corresponding derivative
$$
F_k(u;k)=\int_0^u\frac{k sin^2t dt}{(1-k^2sin^2 t)^{3/2}}
$$
and $A_0, B_0, \vf_0, x^0_A, x^0_B x^0_{\vf}$ are the initial conditions
of the respective phase coordinates.
Now suppose we are given the initial conditions
$\bar u^1,\bar v_1,\bar u^2,\bar v_2$ for the equations of motion of the
reduced system (\ref{rham})
\begin{equation}
\begin{array}{cc}
\displaystyle
\pa_{t_1} u^1=-\frac{\nu^2 u^2}{(v_1 u^2 - J v_2 u^1)^3}
& \pa_{t_2} u^1= v_1 \\[0.5cm]
\displaystyle
\pa_{t_1} u^2=v_2+\frac{J \nu^2 u^1}{(v_1u^2 - J v_2 u^1)^3} &
\pa_{t_2} u^2=J v_2 \\[0.5cm]
\displaystyle
\pa_{t_1} v_1=-\frac{J\nu^2 v_2}{(v_1 u^2 - J v_2 u^1)^3} &
\pa_{t_2} v_1=0 \\[0.5cm]
\displaystyle
\pa_{t_1} v_2=\frac{\nu^2 v_1}{(v_1 u^2 - J v_2 u^1)^3} &
\pa_{t_2} v_2=0.
\end{array}
\label{mot}
\end{equation}
The initial point of the evolution curve (\ref{mot}) is mapped via the
embedding (\ref{emb}) to the point in the extended phase space with the
phase coordinates $\bar x^{\mu},\bar p_{\nu}$\,, such that $\bar
x^1=\bar
p_1=0$ and the gauge fixing condition $\bar x^2=0$ is satisfied. Here we
assume for simplicity that the inequality (\ref{ine}) is also
satisfied.
The gauge fixing condition and the gauge invariant Hamiltonians
(\ref{Cas}) in terms of
the phase coordinates (\ref{AB}),(\ref{ang1}),(\ref{newc}) take the
following form
\begin{equation}
A  x_A cos \vf =  x_{\vf} sin  \vf,
\label{gaug}
\end{equation}\vspace{0.1cm}
\begin{equation}
C_1=\frac{(p_1)^2}{2}+\frac{A^2}{2} \qquad
C_2=J_1\frac{(p_1)^2}{2}+ \frac{B^2}{2}.
\label{Hami}
\end{equation}
Solving the equations of motion with respect to the Hamiltonians
(\ref{Hami}) we get that the momenta (\ref{AB},\ref{ang1}) are conserved
$x^1$ and $p_1$ do not change their zero values and the coordinates
(\ref{newc}) are linear functions of the times $t_1$ and $t_2$,
corresponding to
the Hamiltonians  $C_1$ and $C_2$ (\ref{Hami}) respectively, namely
\begin{equation}
\begin{array}{cc}
\displaystyle
x_0^1=0  &  p^0_1 = 0  \\[0.5cm]
\displaystyle
x^0_{\vf}=\bar x_{\vf} & \vf_0=\bar \vf  \\[0.5cm]
\displaystyle
x^0_A=\bar x_A + \bar A t_1 & A_0=\bar A     \\[0.5cm]
\displaystyle
x^0_B=\bar x_B +\bar B t_2 & B_0=\bar B.
\end{array}
\label{evol}
\end{equation}
Here we suspend the phase space coordinates by $0$ and in order to show
that
they should be considered as the initial conditions for the system of
differential equations (\ref{flow}) because in general a point of the
phase
curve (\ref{evol}) does not satisfy the gauge fixing condition
(\ref{gaug})
but just define the orbit that should intersect the level of the gauge
fixing condition in another point.
Substituting now the phase space coordinates of the curve (\ref{ch})
with the initial data (\ref{evol}) into the gauge fixing condition
(\ref{gaug}) we get the following equation
$$
\bar A (\bar x_A+ \bar A t_1)  = \bar x_{\vf}
\left( \sqrt{\frac{1-k^2 sin^2 \bar \vf}{1-k^2 sin^2\vf(s)}} tg\vf(s) -
\right.
$$
\begin{equation} \left.
-k \sqrt{1-k^2 sin^2 \bar \vf}\, (F_k(\vf(s);k)-F_k(\bar \vf;k))
\right),
\label{resto}
\end{equation}
where $\displaystyle k=\frac{\sqrt{J}\bar A}{\bar B}$\,.
Since the r.h.s. of the equation (\ref{resto}) does not explicitly depend on
the parameter $s$
we can forget about it and treat (\ref{resto}) as the equation, defining
$\vf$ as an implicit function of $t_1$. Really if we
extract the dependence on $\vf$ in the r.h.s. of (\ref{resto}) we get
the function
$$
f(\vf)=\frac{tg\vf}{\sqrt{1-k^2 sin^2 \vf}} -k F_k(\vf;k)
$$
with the
positive derivative
$$
f'(\vf)=\frac{1}{cos^2 \vf \sqrt{1-k^2 sin^2 \vf}} >0.
$$
Thus the intersection of the surface, which consists of the gauge orbits, passing through the points of the
phase curve (\ref{evol}), with the level of the gauge fixing condition
(\ref{gaug}) is given by the following equations
\begin{equation}
\begin{array}{c}
\displaystyle
x_{\vf}(t_1,t_2)=\bar x_{\vf}
\sqrt{\frac{1-k^2 sin^2 \bar \vf}{1-k^2 sin^2\vf(t_1)}} \qquad
x_A(t_1,t_2)=\frac{tg \vf(t_1)}{\bar A} \bar x_{\vf}
\sqrt{\frac{1-k^2 sin^2 \bar \vf}{1-k^2 sin^2\vf(t_1)}} \\[0.5cm]
\displaystyle
x_B(t_1,t_2)=\bar x_B+ \bar B t_2-\frac{\bar x_{\vf}}{\bar B}
\sqrt{1-k^2 sin^2\bar \vf}\,(F(\vf(t_1);k)-F(\bar \vf;k))+\\[0.5cm]
\displaystyle
+
\frac{\bar A\sqrt{J}}{\bar B^2}(F_k(\vf(t_1);k)-F_k(\bar \vf;k))\\[0.5cm]
\displaystyle
A=\bar A \qquad B=\bar B \qquad \vf=\vf(t_1),
\end{array}
\label{expli}
\end{equation}
where $\vf(t_1)$ is the implicit function given by (\ref{resto}).
Using the equations (\ref{expli}) we get the explicit solutions of the
differential equations (\ref{mot}) in the following form
\begin{equation}
\begin{array}{c}
\displaystyle
u^1(t_1,t_2)=\frac{\sqrt{\bar B^2-J\bar A^2sin^2\vf(t_1)}}{\bar B}
x_B(t_1,t_2) \\[0.5cm]
\displaystyle
u^2(t_1,t_2)=x_A(t_1,t_2)sin\vf(t_1)+\frac{x_{\vf}(t_1,t_2)}{\bar A}
\cos \vf(t_1)+\frac{J\bar A}{\bar B}x_B(t_1,t_2)sin\vf(t_1)\\[0.5cm]
\displaystyle
v_1(t_1,t_2)=\sqrt{\bar B^2-J\bar A^2sin^2\vf(t_1)} \qquad
v_2(t_1,t_2)=\bar Asin \vf(t_1),
\end{array}
\label{explic}
\end{equation}
where the initial data in the extended phase space can be obtained with
the help of the following formulas
\begin{equation}
\begin{array}{c}
\displaystyle
\bar A = \sqrt{(\bar v_2)^2+\frac{\nu^2}{(\bar v_1\bar u^2-J\bar v_2
\bar u^1)^2}} \qquad
\bar B=\sqrt{(\bar v_1)^2 + J(\bar v_2)^2}\\[0.5cm]
\displaystyle
\bar \vf = -arctg\frac{\bar v_2(\bar v_1\bar u^2-J\bar v_2
\bar u^1)}{\nu}\\[0.5cm]
\displaystyle
\bar x_A= \bar u^2 sin \bar \vf -
\frac{J\bar A sin^2 \bar \vf}
{\sqrt {\bar B^2-J\bar A^2 sin^2\bar \vf}}\bar u^1 \\[0.5cm]
\displaystyle
\bar x_B=\frac{\bar B \bar u^1}{\sqrt {\bar B^2-J \bar A^2 sin^2
\bar \vf}} \\[0.5cm]
\displaystyle
\bar x_{\vf}= \bar A \bar u^2 cos\bar \vf
-\frac{J\bar A^2 sin\bar \vf cos \bar\vf}
{\sqrt {\bar B^2-J\bar A^2 sin^2 \bar \vf}} \bar u^1.
\end{array}
\end{equation}
\section{Quantization}
In this section we consider the canonical quantization of the presented
system restricting ourselves to the case of the positive $J=J_{32}>0$\,, that
is to the case when the quantum Hamiltonian $\hat{H}_2$ is an elliptic operator.
We realize the Hamiltonians (\ref{rham}) at the quantum level in such a way that the strong
involution relation between them is not destroyed by quantum corrections.
Then, given these commuting Hamiltonians,
we solve the respective eigenvalue problem by requiring that the wave functions
have no branch points.
The surprising thing is that the spectrum of the Hamiltonian $\hat{H}_2$
turns out to be discrete and the coupling constant $\nu^2$ turns out to be
negative.
Our starting point is the classical Hamiltonians
\begin{equation}
H_1=\frac{(v_2)^2}{2J}+\frac{\nu^2}{2J(v_1u^2-v_2u^1)^2}
\qquad
H_2=\frac{(v_1)^2}{2}+\frac{(v_2)^2}{2},
\label{clH}
\end{equation}
which are achieved from (\ref{rham})
by the obvious rescaling
$$
v_1\mapsto v_1 \qquad u^1\mapsto u^1
$$
$$
v_2\mapsto \frac{v_2}{\sqrt{J}} \qquad u^2\mapsto \sqrt{J}u^2.
$$
We perform the quantization of the system (\ref{clH}) in the
momentum representation using the polar coordinates on the
momentum plane
\begin{equation}
v_1=r sin \phi
\qquad
v_2=r cos \phi.
\label{pol}
\end{equation}
In this setting
$$
\hat u_1=i\h(sin\phi \pa_r+\frac{cos\phi}{r}\pa_{\phi}) \qquad
\hat u_2=i\h(cos\phi \pa_r-\frac{sin\phi}{r}\pa_{\phi})
$$
\begin{equation}
\hat v_1=r sin \phi
\qquad
\hat v_2=r cos \phi,
\label{oper}
\end{equation}
and the quantum Hamiltonians take the following form
\begin{equation}
\hat H_1=\frac{r^2 cos^2\phi}{2J}+
\frac{\nu^2}{2J\hat M^2}
\qquad
\hat H_2=\frac{r^2}{2},
\label{quH}
\end{equation}
where the operator $\hat M$ is defined as
$$
\hat M= \hat v_1 \hat u_2 - \hat v_2 \hat u_1=-i\h\pa_{\phi}.
$$
In order to determine the inverse operator to $\hat M^2$
we use the following anzats for the basis wave functions
\begin{equation}
\psi(r,\phi)=R(r)\Phi(\phi)
\label{fact}
\end{equation}
and represent the factor $\Phi(\phi)$ as the
respective Fourier series
\begin{equation}
\Phi(\phi)=\sum_{m=-\infty}^{\infty}\Phi_m e^{im\phi},
\label{F}
\end{equation}
$$
\Phi_m=\frac{1}{2\pi}\int_0^{2\pi}\Phi(\phi)e^{-im\phi}.
$$
Then the operator is naturally defined
on the functions (\ref{fact}) by linearity
if we set that the action of $\hat M^{-2}$
on the eigenfunctions $e^{im\phi}$ of $\hat M$ looks as follows
\begin{equation}
\hat M^{-2}e^{im\phi}=\frac{1}{(\h m)^2} e^{im\phi}.
\label{-1}
\end{equation}
Of course we assume here that the zero mode in the expansion (\ref{F}) is
vanishing $\Phi_0=0$\,.
It is easy to see that thus defined Hamiltonians (\ref{quH}) are commuting operators
and one may pose the respective eigenvalue problem
\begin{equation}
\hat H_1 \psi=E_1\psi
\qquad
\hat H_2\psi= E_2\psi.
\label{prob}
\end{equation}
Using the same anzats (\ref{fact}) for the wave function we see that
this problem (\ref{prob}) reduces to a single
iteration equation for the coefficients $\Phi_m$
\begin{equation}
m^2(\Phi_{m+2}+\Phi_{m-2}-c\Phi_{m})+b\Phi_m=0
\label{iter}
\end{equation}
since the dependence of the wave function
(\ref{fact}) on $r$ is obviously defined by the second equation in (\ref{prob})
$$
R(r)=\de(r-\sqrt{2E_2}).
$$
The iteration equation (\ref{iter}) is equivalent to the following differential equation for the
generating function $\displaystyle \Phi(z)=\sum_{m}\Phi_m z^m$
\begin{equation}
(z\pa_z)^2((z^2+\frac{1}{z^2}-c)\Phi(z))+b\Phi(z)=0,
\label{otz}
\end{equation}
where
$$
c=2(2\frac{JE_1}{E_2}-1)  \qquad b=\frac{2\nu^2}{E_2\h^2}.
$$
In what follows it is more convenient to work with
the unfolded form of this equation (\ref{otz}), namely
\begin{equation}
z^2(z^4-cz^2+1)\Phi''(z)+z(5z^4-cz^2-3)\Phi'(z)+
(4z^4+4+b)\Phi(z)=0.
\label{main}
\end{equation}
It goes with out saying that one can hardly find the solutions for the
equation (\ref{main}) in the explicit form. However, we may use the standard results of the
complex analysis and conclude that the solutions are analytical functions
on ${\bf C}$ besides may be the points where the polynomial $P(z)=z^2(z^4-cz^2+1)$
is vanishing. Hence, we can qualitatively analyze asymptotical
the behaviour of the solutions of (\ref{main}) in the vicinity of the roots
of the polynomial $P(z)=z^2(z^4-cz^2+1)$\,.
And thus, we can derive the properties
of the desired functions $\Phi(\phi)$, obtained by
restricting the solutions to the unit circle $\{z=e^{i\phi}~:~\phi\in {\bf R}\}$\,.
In fact we have to require that the restriction $\Phi(\phi)$ is the function
on the circle. It means that the respective analytic function $\Phi(z)$ have no
branch points in the interior of the unit disc $\{z\in{\bf C}~:~|z|<1\}$
or the factors that arise from these branches should combine into the trivial
factor $1$ if the branch point is not the only one. In addition, we claim that
the function $\Phi(\phi)$ belongs to $L^1(S^1)$\,.
The roots
of the polynomial $P(z)=z^2(z^4-cz^2+1)$ that are worthy of notice
are $z_1=0$ for arbitrary values of the parameters $c$ and $b$ $z_{2,3}=\pm 1$ for
$c=2\,,~b \neq 8$ and $z_{4,5}=\pm i$ for $c=-2$ and $b\neq 8$\,.
It is easy to show that the solutions in the vicinity
of the point $z_1=0$ behaves like $\Phi(z)=z^{\mu}\,,~\mu=2\pm\sqrt{-b}$ if
$b\neq -4$ and behaves analytically if $b=-4$\,. Hence $z_1=0$
is the branch point for our solutions if $b\neq -n^2\,,~n\in {\bf N}$\,.
It turns out that there are no other branch points in the interior of the unit disc
and thus we conclude that the admissible values for $b$ are $-n^2$ with natural numbers $n$\,.
Considering the points $z_{2,3}=\pm 1$ for
$c=2\,,~b \neq 8$ and $z_{4,5}=\pm i$ for $c=-2$ and $b\neq 8$ it is easy to verify that
the solutions of (\ref{main}) behaves like $\Phi(z)=(z-w)^{\beta}$ where
$w=z_{2,3}$ or $w=z_{4,5}$ respectively and $\displaystyle \beta=\frac{-3\pm\sqrt{1-b}}{2}$\,.
Since $\forall~ n \in {\bf N}$
$$
\frac{\sqrt{n^2+1}-3}{2}>-1
$$
we conclude that for an arbitrary $b=-n^2\,,~n\in {\bf N}$ there exists a
solution $\Phi(\phi)\in L^1(S^1)$\,.
Note that in virtue of the relation
$$
b=\frac{2\nu^2}{E_2\h^2}=-n^2
$$
the coupling constant $\nu^2$ turns out to be negative since $E_2$\,,
being the eigenvalue of the elliptic operator $\hat H_2$ is
necessarily positive.
Thus, we get the following
spectrum for the Hamiltonians (\ref{quH})
\begin{equation}
E_1>0 \qquad E_2=-\frac{2\nu^2}{\h^2n^2}, \qquad
n \in {\bf N} \qquad \nu^2<0.
\label{spec}
\end{equation}
The discrete spectrum of the Hamiltonian $\hat H_2$ is not so
surprising because if we consider the exponential degeneration of our
system (\ref{=CM}) and quantize it in the representation of the wave
functions $\psi=\psi(u^2,v_1)$ we note that the momentum $v_1$ defining
the coupling constant of the two-particle Calogero-Moser system is
quantized and thus the respective quantum Hamiltonian $\hat H_2$ also has
a discrete spectrum.
\section{Concluding Remarks}
The main lesson we have learned from the above considerations is that
we may not restrict ourselves to the
Lie group symmetries in constructing the integrable systems via the symplectic
reduction method. Of course if we go far beyond
the Lie group symmetries the explicit description of the gauge orbits cannot be
guaranteed and we cannot be sure that we are able to resolve the equations of
motion via the projection method. Nevertheless we suspect that the
non Lie group symmetries may be useful for constructing the mechanisms of
the symplectic reduction for well known integrable systems.
For example, although the mechanisms of
the symplectic reduction for
the Calogero-Moser systems associated to the root systems of the simple Lie
algebras \cite{OPclass} are known \cite{HM} the projection method for these
system has not been developed. We suppose that the projection method for the
systems can be presented if we construct the mechanisms of
the symplectic reduction in the context of the non Lie group
symmetries.
It should be also very interesting
to perform analogous symplectic reduction starting with the
Odesskii bracket, which generalizes the Sklyanin bracket (\ref{Skl}) to the
case of the arbitrary elliptic $r$-matrix \cite{BD}\,. We suppose that the
projection method for the respective systems should be related
to solving the equations of motion
for some higher dimensional tops \cite{Audin} as is the projection method for our
system related to the equations of motion for the ordinary ${\bf so_3}$ top.
As it has been noted in the introduction the problem of finding the Lax
representation for our system seems to
be rather intriguing. Here we mention that this problem raises the
question of whether it is possible to get the quadratic Sklyanin bracket (\ref{Skl}) via the
reduction from the linear Poisson bracket. As far as we know the latter question is
not answered yet even for the quadratic Poisson brackets, related to the constant triangular
$r$-matrices \cite{LuXu}.
Finally, it is worth mentioning that in our paper we quantize the system starting with
the reduced phase space of the model. However, it would be very interesting to
obtain the same quantum mechanics via the respective quantum reduction,
which may be performed with the help of the BRST approach \cite{Hen}.
We suspect that this quantum reduction should be some how related to
the representations of the quantum Sklyanin algebra \cite{Skl2} as are the
quantum Calogero-Moser systems related to the representation of the
respective Lie algebras \cite{OPquant}.
~\\
{\bf Acknowledgments}. I am grateful to  M.A. Olshanetsky and A.M. Levin for many
useful discussions, suggestions and valuable comments. I also acknowledge M.A. Olshanetsky
for a careful reading of the manuscript and constructive criticisms concerning the
first version of this article.
I am indebted to A. Gorsky and A. Zabrodin for useful remarks.
I would like to express my sincere thanks to S. Lyakhovich for
support and A. Morozov for hospitality and support during my stays
in Moscow. It is my pleasure to thank for hospitality Tamm Theory
Division of the Lebedev Physical Institute where a part of this work
has been done. I also
acknowledge discussions with Yu. Chernyakov, I. Gordeli, A. Kotov, S. Loktev,
D. Talalaev and A. Zotov.  My work is partially supported by RFBR grant
00-02-17-956 and Grant for Support of Scientific Schools  00-15-96557.

\small{

}
\end{document}